\documentclass[
  ,final            
  ]
  {aipproc}

\layoutstyle{8x11single}
\usepackage{caption}

\begin{document}

\title{Charged-Current Neutral Pion production at SciBooNE}


\author{J. Catala-Perez}{
  address={IFIC (U. Valencia/CSIC)}
}

\begin{abstract}

SciBooNE, located in the Booster Neutrino Beam at Fermilab, collected
data from June 2007 to August 2008 to accurately measure muon neutrino
and anti-neutrino cross sections on carbon below 1 GeV neutrino
energy. SciBooNE is studying charged current interactions. Among them,
neutral pion production interactions will be the focus of this poster.
The experimental signature of neutrino-induced neutral pion production
is constituted by two electromagnetic cascades initiated by the
conversion of the $\pi^0$ decay photons, with an additional muon in the
final state for CC processes.

In this poster, I will present how we reconstruct and select
charged-current muon neutrino interactions producing $\pi^0$'s in SciBooNE

\end{abstract}

\maketitle


\section{SciBooNE}

SciBooNE \cite{AguilarArevalo:2006se} is a muon neutrino scattering
experiment located at the BooNE neutrino beam at fermilab. The 0.8 GeV
mean energy neutrino beam is produced with a 8 GeV proton
beam. Protons hit a beryllium target producing charged pions that are
selected and focused using a magnetic horn. The ability to switch the
horn polarity allows to select $\pi^+$ to produce neutrino beam or
$\pi^-$ to produce anti-neutrino beam. Only neutrino beam is currently
used in this analysis.

SciBooNE detector consists in three sub-detectors: the main detector
SciBar, the electromagnetic calorimeter 'EC', and the muon range
detector 'MRD'.
\begin{itemize}
\item
SciBar\cite{Takei:2009zz} is a fully active and fine grained
scintillator detector that consists in 14,336 bars arranged in
vertical and horizontal planes. SciBar is capable to detect all
charged particles and perform dE/dx based particle identification.
\item
The Electron Catcher (EC)\cite{Giganti:2007zz}, is a lead-scintillator
calorimeter consisting in two planes, one vertical and one horizontal,
with a width corresponding to 11 $X_0$.
\item
The MRD\cite{Walding:2007zz}, consists in 12 steel plates sandwiched
between vertical and horizontal planes of scintillator. The MRD has
the capability to stop muons with momentum up to 1.2 GeV. The MRD
detector is used in this analysis to define Charged Current events by
tagging the outgoing muon.
\end{itemize}

The current analysis is covering SciBar contained events, which means
that events with particles other than muons escaping from SciBar
detector are not being considered. EC detector will be introduced in
the analysis in the near future allowing us to use events with particles
escaping from SciBar in the forward direction and reaching the EC.

NEUT \cite{Hayato:2002sd} event generator is used in this
analysis. The Rein-Sehgal model is implemented to simulate Charged
Current resonant pion production with an axial mass $M_A = 1.2$
GeV/$c^2$. All resonances up to 2 GeV are taken into account. However
$\Delta(1232)$ is the resonance that more largely contributes to the
$\pi^0$ production.

\section{CC-$\pi^0$ analysis}

A CC-$\pi^0$ event is defined in this analysis as such event that
contains at least a muon and a neutral pion coming out from the
interaction vertex. This definition includes neutral pions generated
by secondary interactions inside the target nucleus as, for instance,
charge exchanges.

\begin{figure}
  \includegraphics[height=.2\textheight]{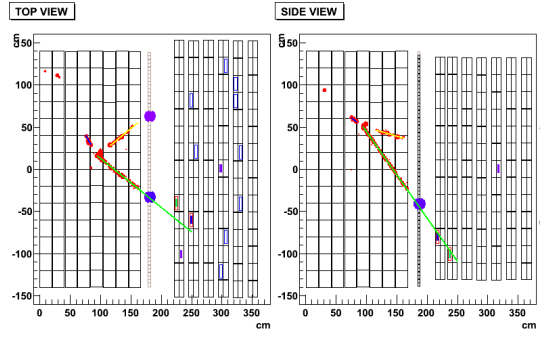}
  \caption{Typical CC-$\pi^0$ event. Muon track in green,
    reconstructed EM showers in yellow and blue.}
  \label{fig:event}
\end{figure}

Though the $\pi^0$ decays almost immediately to two photons, and those
produce EM cascades with an average flight distance of 25 cm,
topologically a CC-$\pi^0$ SciBar contained event contains a muon
reaching the MRD and two or more tracks contained in SciBar (see
fig. \ref{fig:event}). The non-muon tracks are considered gamma
candidates and are used to, at the end, reconstruct the neutral pion.

\subsection{CC-$\pi^0$ selection}

Given the signal definition we can use some event topology and track
property based cuts in order to reduce the background events in the
sample (see table \ref{tab:summary} for summary). The chosen filters
are applied sequentially as follows:
\begin{itemize}
\item
SciBar uses a CC event definition based on the muon tagging using the
MRD\footnote{We are also performing analysis that uses Michel
  electrons to tag the muons that decays in SciBar without reaching
  the MRD.}. Then, the first applied selection is over events that
contains a track reaching the MRD tagged as a muon. Because we don't
expect any other particle to reach the MRD, we also require only one
tagged muon in the event.
\item
Given that we are selecting SciBar contained events, we use a veto
filter to dismiss events with outgoing tracks. The veto filter applies
to events with outgoing tracks either from the upstream or the sides
of the detectors. The veto filter does not apply on tracks pointing to
the EC because those tracks will be fully reconstructed once the EC
information will be used. The veto filter is also useful in order to
remove events with in-going tracks originated in interactions outside
the detector (called DIRT interactions).
\item
As discussed before, we expect events with 3 tracks in SciBar, the
muon and the 2 electromagnetic cascades from the pion decay. We thus
use a filter to meet this topology.
\item
We also use a time based filter in order to avoid cosmic rays and DIRT
generated tracks in our selected events. This filter requires that the
photon candidates should match the muon time with a difference of 20
ns or less.
\item
As commented before, we use the SciBar dE/dx capability in order to
separate minimum ionizing particles as muons or photons\footnote{We
  use 'photon' and 'gamma' in this context to refer the EM cascades
  produced by photon interactions in SciBar.} from protons. Most
protons are rejected using this filter.
\item
Finally, a cut is placed requiring that the photon tracks should be
disconnected from the event vertex taking advantage of the larger photon
flight distance. This cut is particularly useful to reject protons and
charged pions, which track starts always from the event vertex.
\end{itemize}

\begin{table}
\begin{tabular}{lrrrr}
\hline
  & \tablehead{1}{r}{b}{DATA}
  & \tablehead{1}{r}{b}{MC\tablenote{Normalized to CC-Inclusive}}
  & \tablehead{1}{r}{b}{Purity}
  & \tablehead{1}{r}{b}{Efficiency\tablenote{Efficiency relative to
    all events produced in the SciBar fiducial volume.}}   \\
\hline
MRD matched sample & 30161 & 30161 & 6.80\%  & 23.9\% \\
1 muon & 28931 & 27223 & 6.00\%  & 19.2\% \\
veto & 23457 & 24265 & 4.00\% & 11.4\% \\
3 tracks & 912 & 934 & 23.2\% & 2.54\% \\
Time cut & 846 & 903 & 23.8\% & 2.52\% \\
dE/dx & 433 & 447 & 27.5\% & 1.44\% \\
Distance cut & 141 & 137 & 45.5\% & 0.73\% \\
\hline
\end{tabular}
\caption{Event selection summary.}
\label{tab:summary}
\end{table}


\subsection{Preliminary Results}

\begin{figure}
  \includegraphics[height=.2\textheight]{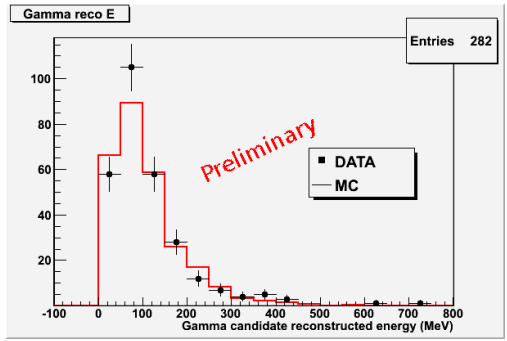}
  \caption{Reconstructed photon energy. MC normalized to CC-Inc.}
  \label{fig:photonE}
\end{figure}

After the above commented cuts, we get reconstructed photons with a
typical energy between 50 and 200 MeV (see
fig. \ref{fig:photonE}). Also, for correctly associated photon
candidates, the energy is reconstructed with 100 MeV resolution and
small bias. The photons are reconstructed at all angles.

\begin{figure}
  \includegraphics[height=.2\textheight]{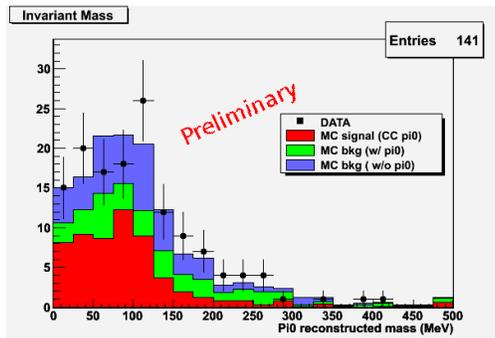}
  \caption{$\pi^0$ reconstructed mass. MC background broken in events
    with neutral pion and events without neutral pion. MC normalized
    to CC-Inc events.}
  \label{fig:mass}
\end{figure}

\begin{figure}
  \begin{minipage}[b]{0.48\linewidth}
    \centering
    \includegraphics[height=.2\textheight]{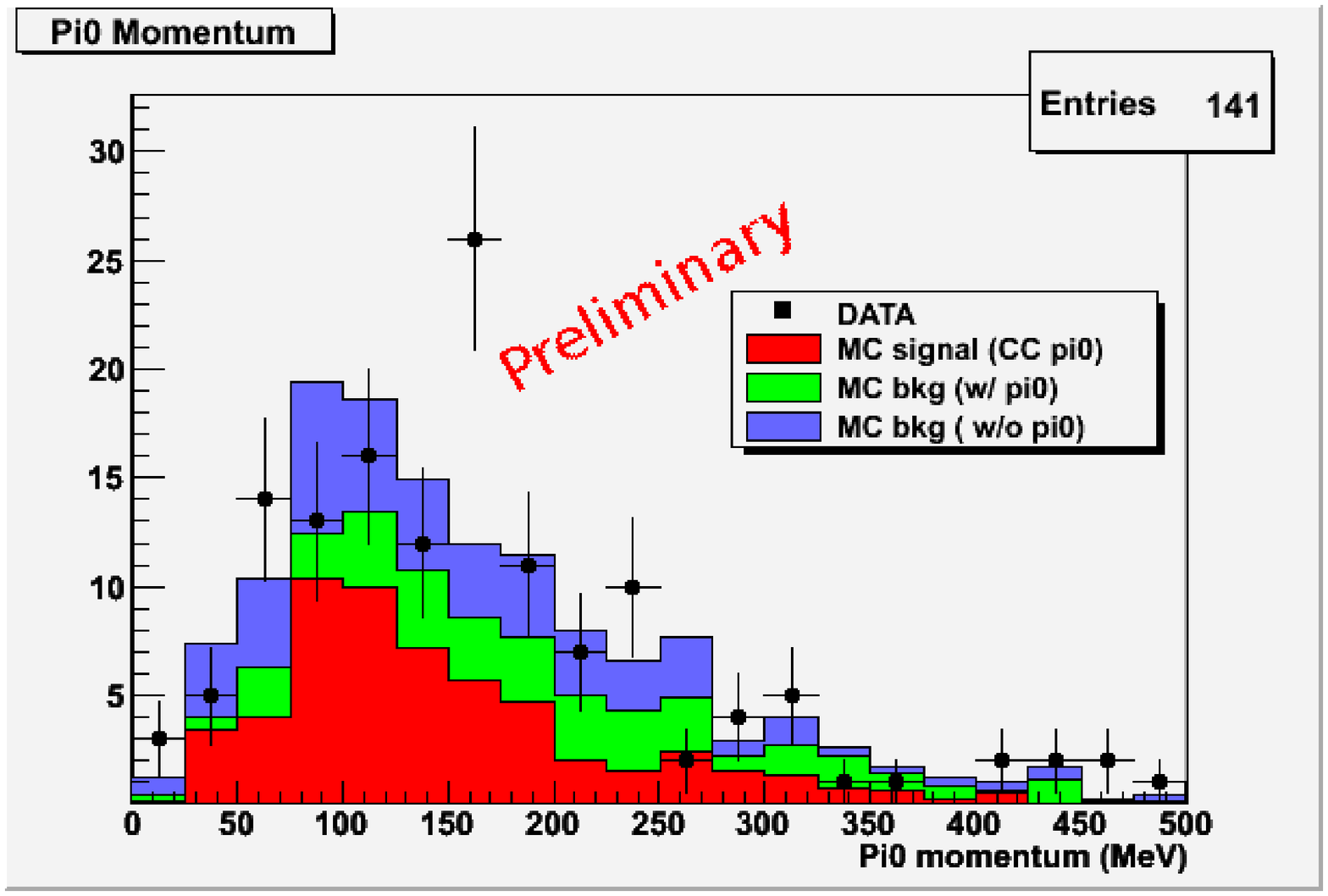}
  \end{minipage}
  
  \begin{minipage}[b]{0.48\linewidth}
    \centering 
    \includegraphics[height=.2\textheight]{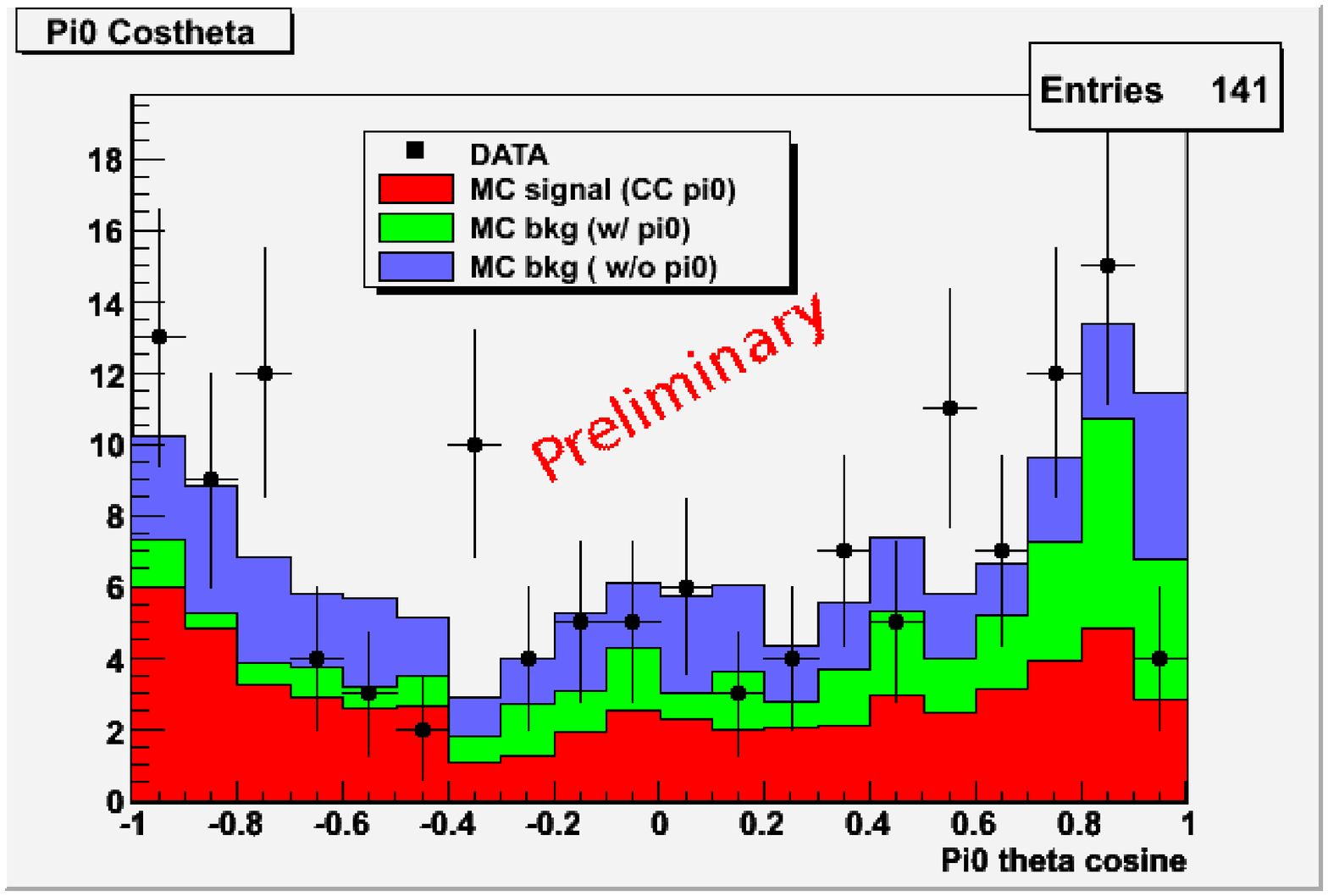}
    \captionof{figure}{\emph{Right plot:} $\pi^0$ reconstructed
      momentum. \emph{Left plot:} Cosine of the angle of the $\pi^0$ with
      respect to the beam.}
    \label{fig:angle}
  \end{minipage}
\end{figure}

Once we have the 2 reconstructed gammas, we are able to reconstruct
also the $\pi^0$ observables. In particular we reconstructed the
invariant mass and also the momentum and angle. As you can see in
fig. \ref{fig:angle} neutral pions are produced at all angles with a
momentum in 50 - 300 MeV/c range. It is also visible a peak in the
invariant mass plot near the $\pi^0$ mass (fig. \ref{fig:mass}).

From the plots we can see that our NEUT-based MC reproduces well the
$\pi^0$ observables.

\section{Updates}
\label{sec:updates}

Since the poster was presented, some reconstruction improvements have
been performed, in this section we are going to discuss them. 

\begin{figure}
  \includegraphics[height=.2\textheight]{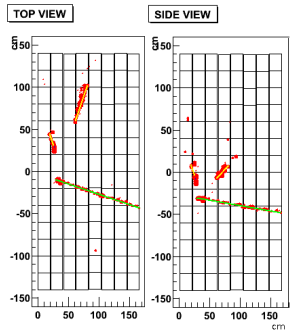}
  \caption{Event reconstructed using the improved track reconstruction
    algorithm. Muon track in green, reconstructed EM showers in
    yellow.}
  \label{fig:sbtcat_event}
\end{figure}

The track reconstruction in SciBar is performed by a cellular
automaton which essentially travels among the beam direction
connecting hits to create tracks. The first reconstruction improvement
was to implement in the code a second run of the automaton but this
time traveling in the transversal direction, that is perpendicular to
the beam, and using the hits that are not associated to any track from
the first processing. In this way we found abut a 10\% more events
containing 3 or more tracks and also we got the ability to reconstruct
tracks at larger angle, close to 90 degrees like in
fig. \ref{fig:sbtcat_event}. This has been an important upgrade given
the low statistics of the analysis mainly due to the lack of events
with 3 or more reconstructed tracks.

A second improvement is to use a new algorithm that improves the
reconstruction performance of the EM cascades. The EM cascades in
SciBar are characterized by disconnected track segments and isolated
hits, making difficult to recover and correctly associate all the
photon visible energy. The new algorithm seeks for those disconnected
track segments and merge them into a single extended track via an
energy-flow algorithm.  It also seeks for hits around the gamma
candidate track in order to add the energy coming from those hits to
the gamma track.

I this way, the photon energy reconstruction is improved and so is the
$\pi^0$ observables. By using the new algorithm we find a narrower
invariant mass peak with less low mass $\pi^0$ than in the
fig. \ref{fig:mass}. Also, the bias in the $\pi^0$ momentum is reduced
and it can be observed an increment of the high momentum pions.

\begin{theacknowledgments}
  The SciBooNE collaboration gratefully acknowledges support from
  various grants and contracts from the Department of Energy (U.S.),
  the National Science Foundation (U.S.), the MEXT (Japan), the INFN
  (Italy) and the Spanish Ministry of Education and Science.
\end{theacknowledgments}

\end{document}